\documentclass[aps, prl, showpacs, twocolumn, 10pt]{revtex4-1}
\usepackage{epsfig}
\usepackage{epstopdf}
\usepackage{amsmath}

\newcommand{\be}{\begin{equation}}
\newcommand{\ee}{\end{equation}}
\newcommand{\bea}{\begin{eqnarray}}
\newcommand{\eea}{\end{eqnarray}}
\newcommand{\edot}{\dot{\epsilon}}
\newcommand{\edotbar}{\overline{\dot{\epsilon}}}
\newcommand{\ebar}{\overline{\epsilon}}

\newcommand{\bw}{\begin{widetext}}
\newcommand{\ew}{\end{widetext}}

\newcommand{\tensile}{\sigma_{\rm E}}
\newcommand{\tw}{t_{\rm w}}

\newcommand{\Sig}{\mathbf{\Sigma}}

\newcommand{\vecv}[1]{\mathbf{{#1}}}
\newcommand{\tens}[1]{\mathbf{{#1}}}
\newcommand{\nablu}{{\bf \nabla}}

\begin{document}
\title{Age dependent modes of extensional necking instability in soft glassy
  materials}

\author{D. M. Hoyle}
\affiliation{Department of Physics, Durham University, Science Laboratories, South Road, Durham DH1 3LE, United Kingdom}

\author{S. M. Fielding}
\affiliation{Department of Physics, Durham University, Science Laboratories, South Road, Durham DH1 3LE, United Kingdom}

\date{\today}

\begin{abstract}

  We study the instability to necking of an initially cylindrical
  filament of soft glassy material subject to extensional stretching.
  By numerical simulation of the soft glassy rheology model and a
  simplified fluidity model, and by analytical predictions within a
  highly generic toy description, we show that the mode of instability
  is set by the age of the sample relative to the inverse of the
  applied extensional strain rate.  Young samples neck gradually via a
  liquid-like mode, the onset of which is determined by both the
  elastic loading and plastic relaxation terms in the stress
  constitutive equation. Older samples fail at smaller draw ratios via
  a more rapid mode, the onset of which is determined only by the
  solid-like elastic loading terms (though plastic effects arise
  later, once appreciable necking develops).  We show this solid-like
  mode to be the counterpart, for elastoplastic materials, of the
  Consid\`ere mode of necking in strain-rate-independent solids.

\end{abstract}


\maketitle

Many soft materials, including foams, emulsions, microgels and
colloids comprise disordering packings of mesoscopic substructures:
foam bubbles, emulsion droplets, {\it etc}.  At high volume fractions,
the local rearrangement dynamics of these are impeded by large energy
barriers and show a glassy slowing down. This underpins many universal
features in the rheological (deformation and flow) properties of these
``soft glassy materials'' (SGMs).  Particularly striking is the
phenomenon of rheological ageing, in which an initially liquid-like
sample slowly evolves towards an ever more solid-like state as a
function of the time since it was prepared.  In the last decade,
significant progress has been made in understanding the role of ageing
in the shear flow of SGMs~\cite{fielding-jr-44-323-2000}.  Similar
phenomena have been explored in polymeric~\cite{Struik78} and
metallic~\cite{chen80} glasses, with many unifying features across all
these amorphous, elastoplastic materials.

\begin{figure}[b]
  \includegraphics[width=0.35\textwidth ]{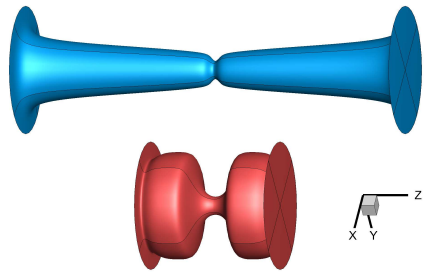}
  \caption{Necked profiles. Extension rate shown by arrow in
    Fig.~\ref{fig:fluidity0}. Top: young sample, $\tw=10^2$.
    Bottom: old, $\tw=10^4$.  ~\label{fig:profiles}}
\end{figure}

Much less is understood about the response of these materials to
extensional deformations, which are important to industrial processes
in fibre spinning, ink-jetting, porous media, and the peeling and tack
of surfaces bonded by adhesives.  In the standard experimental test,
an initially near uniform cylindrical (or rectangular) sample is
steadily drawn out in length, with the aim of measuring the tensile
stress as a function of strain and strain rate.  Ubiquitously
observed, however, is an instability to neck formation: the sample
thins more quickly in the middle than at its endpoints
(Fig.~\ref{fig:profiles}) and eventually fails.  This has been
observed in emulsions~\cite{Huisman2012, Niedzwiedz2010, Derkach2009,
  Miller2009,Coussot2005}, laponite suspensions,~\cite{Shaukat2010,
  Shaukat2012,Coussot2005}, foams~\cite{Arciniaga2011, Kuo2012,
  Kuo2013}, polymer glasses~\cite{lee09,Rottler2009}, simulations of
shear transformation zone models~\cite{Eastgate2003,Rycroft2012}, and
in shear thickening colloids~\cite{Zimoch2013,
  Roche2011,Smith2010,chellamuthu09} (though our focus here is on
shear thinning SGMs).

A hallmark of all these elastoplastic materials is that their
deformation properties depend strongly on the {\em rate} at which
strain is applied, particularly when ageing is present. For rate-{\em
  independent} materials, the onset of necking was predicted by
Consid\`ere in 1885~\cite{Considere} to coincide with a regime of
declining tensile force as a function of strain.  But despite the
accumulating body of observations described above, counterpart
criteria for necking in elastoplastic materials remain lacking (though
for an insightful early study of rate dependence,
see~\cite{hutchinson77}).  The contribution of this Letter is to
provide such criteria, in the form of  general analytical results
supported by numerical simulation of two widely used models of soft
glasses. In this way, we argue these new criteria to apply universally
across all ageing elastoplastic materials and so to have the same,
highly general status as the Consid\`ere criterion for
rate-independent solids.

Importantly, we find the mode of instability to necking to be
determined by the age $\tw$ of the sample, relative to the inverse of
the applied extensional (Hencky) strain rate $\edotbar$: young samples
($\edotbar\tw\ll 1$) fail by a gradual, liquid-like mode, whereas old
samples ($\edotbar\tw\gg 1$) show a fast, solid-like failure mode. We
further show this solid-like mode to be the counterpart for
elastoplastic materials of the Consid\`ere mode in rate-independent
solids.  In this way, crucially, 
the physics of a material's failure is predicted to be governed by a
switch between two qualitatively distinct modes of instability,
governed simply by the sample age~\cite{noteBouchbinder}.

We consider incompressible, inertialess deformations for which the
velocity and stress fields in the material, $\textbf{v}(\textbf{r},t)$
and $\textbf{T}(\textbf{r},t)$, obey standard conditions of mass
balance, $\nabla.\textbf{v}=0$, and force balance, $\nabla.\textbf{T} =
0$.  The total stress $\textbf{T}= \Sig + 2\eta\tens{D}
-p\tens{I}$ comprises an elastoplastic contribution $\Sig$ from
the mesoscopic substructures, a Newtonian solvent contribution of viscosity
$\eta$, and an isotropic pressure. Here $K_{\alpha\beta} =
\partial_\beta v_\alpha$ and $\tens{D}=\tfrac{1}{2}(\tens{K}
+\tens{K}^T)$. 

For the dynamics of the elastoplastic stress $\Sig$ we adopt the soft
glassy rheology (SGR) model~\cite{sollich-prl-78-2020-1997}. This
considers an ensemble of elements, each corresponding to a local
mesoscopic region of material.  Under an imposed deformation, each
element experiences a buildup of local elastic stress, intermittently
released by plastic relaxation events.  The treatment of tensorial
stresses~\cite{cates-jr-48-193-2004} within SGR was inspired by the
Doi-Ohta model of dense emulsions~\cite{Doi1991}, and considers a
local density function $f(\vecv{n})$ for the area (per unit volume) of
droplet interfaces normal to $\vecv{n}$, with a spherical
normalisation $Q=\int d\vecv{n} f(\vecv{n})$ and stress $\Sig = G\int
d\vecv{n} (\vecv{n}\vecv{n}-\tfrac{1}{3}\tens{I}) f(\vecv{n})$.
The constant modulus $G=1$ sets our stress scale. The buildup of
elastic stress in any element during deformation obeys~\cite{Doi1991}
\bea
(\partial_t +\vecv{v}.\nablu)\Sig &=& \Sig.\tens{K} +\tens{K}^T.\Sig  +\frac{2}{3}Q\tens{D} -\Sig:\tens{K}\left(\frac{2}{3}\tens{I}+\frac{\Sig}{Q}\right),\nonumber\\
(\partial_t +\vecv{v}.\nablu) Q&=&\tens{K}:\Sig.\label{eqn:DO1}
\eea

Relaxation of stress by local plastic yielding events is modelled as
hopping of the elements over strain-modulated energy barriers,
governed by a noise temperature $x$. Upon yielding, any element resets
its local stress to zero and selects its new energy barrier at random
from an exponential distribution.  This distribution confers a broad
spectrum of yielding times $P(\tau)$, resulting in a glass phase with
a yield stress for $x<1$.  Full details of the model in its
original, spatially uniform form are in
Ref.~\cite{sollich-prl-78-2020-1997,cates-jr-48-193-2004}, and in its
adaptation to spatially non-uniform shear flows in
Ref.~\cite{Fielding2009}.  The counterpart adaptation for non-uniform
extension is summarised in~\cite{SI}.

Because the SGR model is numerically rather cumbersome, we shall also
present results for a simplified fluidity model in which SGR's full
spectrum of yield times $P(\tau)$ is replaced by a single
characteristic relaxation timescale $\tau$. The RHS of
Eqs.~\ref{eqn:DO1} then acquire relaxation terms:
$-\frac{1}{\tau}Q\Sig$ and $-\frac{1}{\tau}\mu Q^2$ for the $\Sig$ and
$Q$ dynamics respectively. Here $\mu$ is a phenomenological parameter
with $0\le \mu\le 1$~\cite{Doi1991}.  A standard ``fluidity'' model
for the dynamics of $\tau$,
\be 
(\partial_t +\vecv{v}.\nablu) \tau = 1 - \sqrt{2\tens{D}:\tens{D}} \left( \tau - \tau_0 \right),\label{eqn:DO3} 
\ee
then captures the two essential ingredients of the SGR model: (i)
ageing without flow, in which the relaxation time increases with the
time since sample preparation, $\tau\sim t$; and (ii) rejuvenation by
flow, which restores a steady state with
$\tau=\tau_0+1/\sqrt{2\rm{Tr}\tens{D}:\tens{D}}$.  We choose units in
which the microscopic time $\tau_0=1$.


We consider an initially cylindrical sample of length $L_0$ and radius
$R_0$, freshly prepared at time $t=0$ in a fluidized state with
$\tau=\tau_0$ and $\tens{\Sigma}=\tens{0}$. It is then left to age
undisturbed during a time $\tw\gg \tau_0$, before being steadily drawn
out such that for times $t > \tw$ the length increases as
$\dot{L}=\edotbar(t) L$, with the cross sectional area correspondingly
thinning. We present results below for two common protocols: (a)
constant rate of Hencky strain $\edotbar$, corresponding to
$L=L_0\exp\left[\edotbar (t-\tw)\right]$, and (b) constant rate
$\dot{L}$ of length increase, corresponding to a progressively
declining strain rate $\edotbar = (1/\edotbar_0+ t-\tw)^{-1}$.

For convenience, we solve the models within an one-dimensional (1D)
approximation~\cite{Denn1975, Olagunju1999, SI}, in which the
wavelengths of any variations that develop along the filament are
assumed long compared to the radius.  This standard assumption~\cite{
  Yao1998, Tembely2012, Vadillo2012} has been shown to perform
surprisingly well even some way into the regime where the wavelengths
become comparable to the radius. It allows the neglect of any radial
dependencies, such that the deformation of a filament extended in the
$z$ direction is characterised simply by its cross-sectional area
$A(z, t)$ and an area averaged $z-$component of velocity $v(z, t)$.

A choice must then be made for how to model the no-slip condition
where the sample ends meet the experimental endplates.  A good
approximation~\cite{Stokes2000, Clasen2006, Szabo2012} for initial
aspect ratios $\Lambda = L_0/R_0 \geq 1$ is to invoke a divergent
viscosity over a small region of the filament near each plate, acting
to  pin the fluid to the plates.  (Equivalently, this region can be
thought of as {\it part} of the plates.) We adopt this
assumption~\cite{SI}, and have checked that the physics reported is
robust to it by also performing simulations (not shown) with periodic
boundary conditions (corresponding to a stretched torus, without
endplates).

\begin{figure}[t]
\includegraphics[width=0.425\textwidth ]{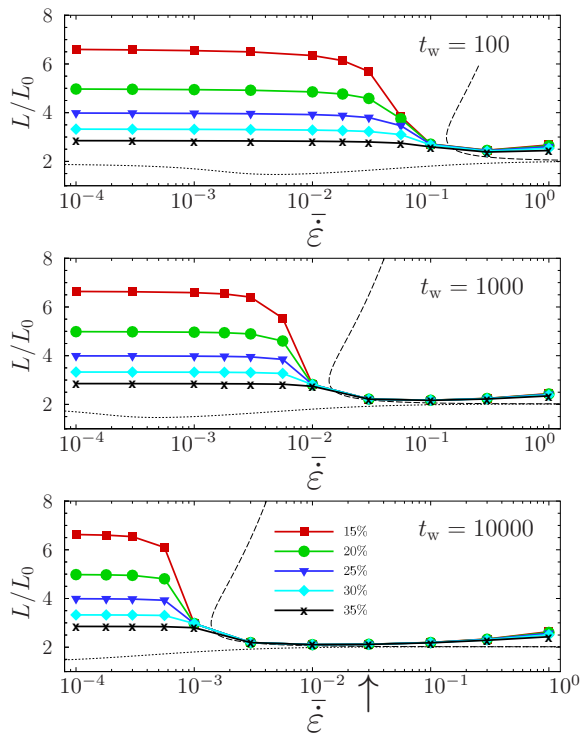}
\caption{Fluidity model in protocol a). Solid lines: draw ratio at
  which the radius of the filament at its thinnest point has fallen to
   35\%, 30\% $\cdots$ 15\% of the initial filament radius (curves
  upwards). Data are shown for three different initial sample ages
  (panels downwards). Initial aspect ratio $\Lambda=2$. Dotted lines:
  Consid\`ere criterion~\cite{Considere}.  Dashed lines: counterpart
  of the Consid\`ere criterion for elastoplastic materials.
  Parameters: $\mu=0.1$ and $\eta = 0.01$. Arrow shows $\edotbar$ for
  Fig.~\ref{fig:profiles}.
  \label{fig:fluidity0}}
\end{figure}

We neglect surface tension, restricting to the development of a neck
in a highly elastoplastic filament in which bulk stresses dominate.
Clearly, surface tension must ultimately become important in the very
final stages of breakup, once the radius becomes small~\cite{noteST}.  However our
focus is not on the details of that final pinch-off, but on the
time at which necking first becomes appreciable, which we define as the centrepoint-radius having fallen to 15\% of the initial radius.

The basic phenomenon that we seek to explain is shown in
Fig.~\ref{fig:profiles}. This displays the necked profiles of two
filaments: one that has been drawn relatively slowly compared to the
inverse age and another that has been drawn more quickly.  As can be
seen, the slowly drawn filament survives to a relatively longer draw
ratio before approaching failure, and displays a gradual necking
profile.  In contrast, the rapidly drawn filament fails sooner and
with a more pronounced, cusp-like profile.

\begin{figure}[t]
\includegraphics[width=0.425\textwidth ]{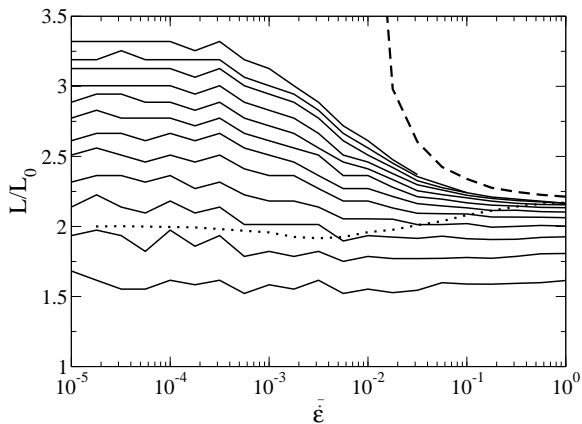}
\caption{SGR model in protocol a).  Solid contour lines upward show
  where initial area perturbations have increased in amplitude by
  successive powers of  $1.4$. The initial area profile was seeded such that $A = 1 + \xi \cos ( 2 \pi u )$, where $\xi = 3\cdot 10^{-4}$. Dotted line: Consid\`ere criterion. Dashed
  line: counterpart criterion for elastoplastic material.  Parameters:
  $x=0.3$, $\lambda=1.5$, $\tw=10^2$. ~\label{fig:SGR}}
\end{figure}

That basic observation is quantified for the fluidity model in
Fig.~\ref{fig:fluidity0}, for protocol (a) in which the Hencky strain
rate $\edotbar$ is held constant and the sample ends separate
exponentially. The symbols show the draw ratio $L(t)/L_0$ at which
appreciable necking becomes apparent, versus that applied strain rate.
As can be seen, old samples ($\edotbar\tw \gg 1$) fail at a modest
draw ratio $L/L_0$, while young samples ($\edotbar\tw \ll 1$) survive
to a larger $L/L_0$. Also shown (dotted line) is the Consid\`ere
criterion at which the tensile force first starts declining with
strain. This would signal necking onset in a rate-independent solid
but evidently performs badly here for young samples $\edotbar\tw < O(1)$.
We return below to discuss this in the context of an alternative
criterion shown by the dashed lines. Fig.~\ref{fig:SGR} confirms the
same behaviour in the SGR model.  The same behaviour is also
seen~\cite{SI} in protocol (b), which has a constant drawing rate
$\dot{L}=\edotbar_0L_0$ and progressively decreasing strain rate
$\edotbar =(1/\edotbar_0+ t-\tw)^{-1}$.

To allow analytical insight, let us consider now a highly simplified,
toy description of an elastoplastic filament in extensional
deformation.  The relevant dynamical quantities are the area profile
$A(z,t)$, the $z$-component of velocity $v(z,t)$, the strain rate
field $\edot(z,t)=\partial_z v(z,t)$, with the overall applied strain
rate $\edotbar(t)=\int_0^L dz \edot(z,t)/L$, the elastoplastic stress
$GZ(z,t)$, which we write as a constant modulus $G$ times a
strain-like variable $Z$, the total tensile stress
$\tensile(z,t)=GZ+3\eta\edot$, and force $F(t)=\tensile A$.

These obey the conditions of mass and force balance:
\bea
\partial_tA+v\partial_z A&=&-\edot A\nonumber\\
0&=&\partial_z(A\tensile).\label{eqn:toy1}
\eea
We then choose for $Z$ the simplest possible dynamics that combines
elastic loading ($f$) and plastic relaxation ($g$):
\be
\partial_tZ+v\partial_zZ=\edot f(Z)-\frac{1}{\tau}g(Z).\label{eqn:toy2}
\ee
We intentionally leave the forms of $f$ and $g$ unspecified, in order
to derive below general instability criteria that do not depend on
particular choices.  For simplicity we do not include in this toy
description explicit dynamics for $\tau$, but below comment on
predictions when $\edotbar\tau\gg 1$ and $\edotbar\tau\ll 1$, thereby
inserting ageing $\tau\sim\tw$ ``by hand''.

Within this toy model, we consider an initially near uniform
cylindrical filament and express its state as the sum of a
time-dependent uniform base state, corresponding to a perfect cylinder
being stretched, plus small amplitude heterogeneous perturbations.
Accordingly we write $a(u,t)=\bar{a}+\delta a_q\exp(iqu)$,
$Z=\bar{Z}(t)+\delta Z_q\exp(iqu)$, $\edot(u,t)=\edotbar(t)+\delta
\edot_q\exp(iqu)$, choosing for convenience to work in the
coextending, cothinning frame by defining transformed length and area
variables $u=z\exp(-\ebar)$ and $a=A\exp(\ebar)$.  We then perform a
linear stability analysis to determine the time at which these
heterogeneous perturbations start to grow, corresponding to the onset
of necking. Substituting into Eqns.~\ref{eqn:toy1} and expanding in
powers of the perturbation amplitude gives at first order
\be
\label{eqn:linear}
\partial_t\left(\begin{array}{c}\delta a\\\delta Z\end{array}\right)_q=\tens{M}_q(t)\cdot\left(\begin{array}{c}\delta a\\\delta Z\end{array}\right)_q,
\ee
in which the stability matrix $\tens{M}$ has inherited the time-dependence
of the uniform base state. 

At least one eigenvalue of $\tens{M}$ being positive at any time $t$
then gives a strong indication that the heterogeneous perturbations
will be growing at that time, corresponding to the development of
necking.  (Direct integration of the linearised equations confirms
this, and also agrees with the early-time growth of a neck in a full
nonlinear solution.) In fact it is straightforward to show that
$\tens{M}_q$ has two distinct modes of instability. One of these,
which we call mode 1, has an eigenvalue of order $\edotbar$ that is
positive when the tensile stress of the underlying base state obeys
$\edotbar f'-\tfrac{1}{\tau}g'=\ddot{\tensile}/\dot{\tensile}<0$. It
is liquid-like in the sense that its onset is determined by an
interplay between elastic loading (specified by $f$) and plastic
relaxation (via $g$).  The other (mode 2) has a much larger eigenvalue
of order $G/\eta$ that is positive when the base state $\tensile-Gf >
0$: its onset condition is set only by the solid-like elastic loading
term $f$, independent of $g$~\cite{noteEC}.

These analytics explain our numerical results in the SGR and fluidity
models as follows. We indeed see a mode directly analogous to mode 1,
the onset of which involves both the elastic loading and plastic
relaxation dynamics, and which involves significant plastic relaxation
along the entire filament.  Its quantitative onset criterion is
however modified compared with the toy model, due to the higher
dynamical dimensionality of the full models: it is actually unstable
for all strain rates and at all times during stretching. Any small
perturbations to an initially cylindrical profile therefore start
slowly growing as soon as stretching starts, for all strain rates.  It
is this mode that eventually leads to failure at relatively large
$L/L_0$ for young samples $\edotbar\tw\ll 1$ in
Figs.~\ref{fig:fluidity0} and~\ref{fig:SGR}.

For old samples, the solid-like mode 2 intervenes to cause failure at
smaller strains.  Its onset criterion $\tensile-Gf>0$ transcribes
(once cast into the forms of (\ref{eqn:EC},\ref{eqn:ECb}) below)
unmodified to the fluidity and SGR models (and indeed any model,
however complex). It is shown by the dashed line in
Figs.~\ref{fig:fluidity0} and~\ref{fig:SGR}, and is only satisfied if
$\edotbar\tau \gg 1$, {\it i.e.} in old samples for which
$\edotbar\tw\gg 1$.  It is furthermore only met a finite time after
the start of stretching.  Once unstable, however, its much larger
eigenvalue $O(G/\eta)$ gives much more rapid necking and the sample
fails very shortly after instability onset, at only modest draw
ratios.  (This mode would in principle restabilise at higher strains,
as seen in Fig.~\ref{fig:fluidity0} by the doubling-back of the dashed
line, but that is irrelevant because the filament will have failed by
then.)

How do we understand this instability criterion, $\tensile-Gf > 0$ of
mode 2?  Noting that the tensile force $F=A\tensile$, we have (for the
underlying base state)
\be 
\label{eqn:C}
\frac{\partial F}{\partial \epsilon}=A\left(-\tensile +
  \frac{\partial \tensile}{\partial\epsilon}\right)=A\left(-\tensile+Gf-G\frac{g}{\edot\tau}\right).  
\ee
This follows from Eqns.~\ref{eqn:toy1} and~\ref{eqn:toy2}, neglecting
terms $O(\eta/G)$.  Were the Consid\`ere criterion to apply directly,
that would give instability for $\tensile-Gf+Gg/\edot\tau >0$.
However the dotted lines in Figs.~\ref{fig:fluidity0} and ~\ref{fig:SGR}
show this criterion to perform poorly for
young samples $\edotbar\tw < O(1)$.  We can, though, define a modified
derivative,
\be
\label{eqn:EC}
\frac{\partial F}{\partial \epsilon}|_{\rm elastic}=A\left(-\tensile+Gf\right), 
\ee
in which the plastic relaxation term is artificially switched off over
the strain increment in question.  We thereby recognise the onset of
mode 2 instability as
\be
\label{eqn:ECb}
\frac{\partial F}{\partial \epsilon}|_{\rm elastic} < 0.
\ee
It is this criterion that is marked as a dashed line in
Figs.~\ref{fig:fluidity0} and~\ref{fig:SGR}. It performs much better
for these ageing materials, over the full range of $\edotbar\tw$, than
the original Consid\`ere criterion.  It coincides with Consid\`ere
only for $\edotbar\tw\gg 1$: in this regime, the material behaves
essentially as a nonlinear elastic solid (with $\edot f\gg
\tfrac{g}{\tau}$), at least until necking first arises.  (Once
appreciable necking occurs it causes plastic flow in the thinning,
central region of the filament.)  Accordingly, we propose
(\ref{eqn:ECb}) as the counterpart for rate-dependent elastoplastic
materials of the Consid\`ere criterion for rate-independent solids.

Finally, we have checked that our results are robust to reasonable
variations in the values of the initial aspect ratio $\Lambda$ and the
Newtonian viscosity $\eta$; and that they apply robustly over large
regions of the space of the phenomenological parameters $\mu$
(fluidity) and $\lambda,x$ (SGR, within its glass phase $x<1$). See
the SI for details~\cite{SI}.

To summarise, we have shown the instability to necking of a filament
of soft glassy material to proceed by one of two possible modes.
Young samples neck gradually via a mode informed by both elastic
loading and plastic relaxation. Old samples fail more suddenly via a
mode in which onset is informed only by elastic loading, and which is
the counterpart for elastoplastic materials of the Consid\`ere mode of
rate-independent solids.  A particularly appealing feature of the
physical picture presented here is this crossover between two distinct
modes of instability, determined simply by the age of the sample
relative to the inverse stretching rate. Having shown
this numerically in two widely used models of soft glasses, as well as
analytically, we argue these predictions to apply generically to all
ageing elastoplastic materials.  Indeed they may apply even more
generally still, to pasty materials with long relaxation timescales
$\tau$ but much weaker ageing effects, now set by the value of the
inverse strain rate relative to $\tau$. It remains an open challenge
to understand how, within SGR, these ductile failure modes cross over
to brittle cracking at even higher strain rates, for samples with
notch-like initial imperfections~\cite{falk99,Rycroft2012a}.

{\em Acknowledgements ---} The research leading to these results has
received funding from the European Research Council under the European
Union's Seventh Framework Programme (FP7/2007-2013) / ERC grant
agreement number 279365.


\begin{thebibliography}{50}
\bibitem[38]{noteST}  The contributions to force balance of the viscoelastic stresses and surface tension are respectively $G \partial_z \Sigma$ and (at leading order) $\gamma \partial_z R^{-1}$ where gamma is the coefficient of surface tension and R the radius of the filament.
\bibitem[39]{noteEC} Constitutive models that are too
  oversimplified to address the physics soft glasses, such as the
  Oldroyd B model, have an elastic loading term of trivially simple
  form and this mode 2 instability criterion is never met.
\end{thebibliography}

 \newpage 




  \begin{center}
    \textbf{Supplemental Material for: \\ \textit{Age dependent modes of extensional necking instability in soft glassy
	materials}}
  \end{center}

%
  \setcounter{equation}{0}
  \setcounter{figure}{0}
  \setcounter{table}{0}
  \setcounter{page}{1}
  \makeatletter
  \renewcommand{\theequation}{S\arabic{equation}}
  \renewcommand{\thefigure}{S\arabic{figure}}
  \renewcommand{\bibnumfmt}[1]{[S#1]}
  \renewcommand{\citenumfont}[1]{S#1}
  \appendix


\section*{ Fluidity model }

In this section, we detail a one-dimensional
approximation~\cite{Denn1975_SI, Olagunju1999_SI} in which the wavelengths
of any variations along the filament are assumed long compared to the
filament's radius. We also describe a transformation to the frame of
reference that coextends and cothins with the overall average
deformation of the filament.  We then discuss the implementation of
the no-slip condition where the filament meets the experimental
endplates. 

\phantom{a} 
\subsection*{One dimensional approximation}

We adopt a one dimensional approximation~\cite{Denn1975_SI, Olagunju1999_SI}
that allows us to neglect any dependencies in the radial direction and
consider only those in the coordinate $z$ along the length of the
filament. The cross sectional area is denoted $A(z,t)$ and the
area-averaged local $z-$component of velocity is denoted $v(z,t)$. The
local elongation rate at any point along the filament is
$\dot\epsilon =\partial_z v$.

The continuity equation then reads
\begin{equation}
\partial_tA + v\partial_zA = - A\dot\epsilon.
\end{equation} 
The force balance condition imposes that the tensile force in the
filament is independent of $z$,
\begin{equation}
\partial_zF = 0.
\end{equation} 
The force $F$ is the product of the area of the filament multiplied by
the tensile stress:

\begin{equation}
F(t) = A(\Sigma +3\eta\dot\epsilon),
\end{equation} 
where the tensile stress comprises additive viscoelastic and Newtonian
contributions.  Here the scalar viscoelastic stress difference
$\Sigma=\Sigma_{zz}-\Sigma_{xx}$, where $x$ denotes a radial
coordinate. We then write $\Sigma=G\tilde{\Sigma}$ where the constant
modulus $G=1$ in our units, defining a strain like variable
$\tilde{\Sigma}$, then drop the tilde for clarity.

The constitutive equations for the viscoelastic stress then read
\begin{eqnarray}
 \partial_t\Sigma + v\partial_z\Sigma  &=& \dot\epsilon\left(Q - \frac{\Sigma^2}{Q}-\Sigma\right)-\frac{1}{\tau}Q\Sigma,\\
 \partial_t Q + v\partial_z Q &=& \dot\epsilon \Sigma - \frac{1}{\tau}\mu Q^2.
\end{eqnarray}
The equation of motion for the relaxation time $\tau$ finally becomes
\begin{equation}
\partial_t\tau  +v\partial_z\tau = 1-(\tau-\tau_0)\sqrt{3}|\dot\epsilon|. \label{suptau}
\end{equation}

\subsection*{Coordinate transformation}

In this section, we describe an affine coordinate transformation to
the co-extending and co-thinning frame. This removes the increase in
sample length, which in the laboratory frame occurs as an exponential
function of the spatially averaged Hencky strain
${\overline\epsilon}(t)$ applied to the filament. It also removes
the corresponding exponential decrease in the filament's area.  It
renders numerical solution of the model much more straightforward, by
removing the need for an adaptive mesh to address the changing sample
length in the laboratory frame.

Accordingly, old variables $z,v,\dot\epsilon,A,F$ are transformed into
new variables $u,w,\dot\zeta,a,f$ following:
\begin{eqnarray}
z &=& u \exp[{\overline\epsilon}(t)],\\
v &=& [\dot{\overline\epsilon}(t)u + w(u,t)]\exp[{\overline\epsilon}(t)],\\
\dot\epsilon &=& \dot{\overline\epsilon}(t) + \dot\zeta(u,t),\\
A &=& A(0) a(u,t)\exp[-{\overline\epsilon}(t)],\\
F &=& A(0) f.
\end{eqnarray}
The other variables, $\Sigma, Q$ and $\tau$, are not transformed, but
are now expressed as functions of the transformed space variable $u$
and time $t$. The transformed equations of motion are then
\begin{eqnarray}
f &=& \exp[-\overline\epsilon(t)]a[\Sigma + 3\eta (\dot{\overline\epsilon}+\dot\zeta)],\label{scale1}\\
 \partial_t a + w\partial_u a &=&-\dot\zeta a, \label{scale2}\\
 \partial_t\Sigma + w\partial_u\Sigma &=& (\dot{\overline\epsilon}+\dot\zeta) 
\left(Q - \frac{\Sigma^2}{Q}-\Sigma\right)-\frac{1}{\tau}Q\Sigma,\\
 \partial_t Q + w\partial_u Q &=& (\dot{\overline\epsilon}+\dot\zeta) \Sigma -\frac{1}{\tau}\mu Q^2,\\
 \partial_t\tau + w\partial_z\tau &=& 1-(\tau-\tau_0)\sqrt{3}|\dot{\overline\epsilon}+\dot\zeta|. \label{scale7}
\end{eqnarray}
To solve these equations numerically, $a, \Sigma, Q$ and $\tau$ are
updated at each time step using Eqns.~\ref{scale2} to~\ref{scale7}.
The updated values are then substituted into Eqn.~\ref{scale1}. An
appropriate integral of Eqn.~\ref{scale1} along the entire filament
determines $f$, which is finally substituted back into
Eqn.~\ref{scale1} to determine the strain rate profile
$\edotbar+\dot\zeta(u,t)$ as a function of distance along the
filament.

We choose our unit of length such that the initial length of the
sample is unity. The sample domain is then $0<u<1$ for all times.  The
initial cross sectional area $A(0)$ defines the undeformed starting
shape of the sample, with $a(u,0) = 1$ for all $u$.

\subsection*{Endplates}

The no-slip boundary condition where the ends of the sample meet the
experimental endplates is imposed following Refs.~\cite{Stokes2000_SI,
  Clasen2006_SI, Szabo2012_SI}, by invoking a divergent viscosity in a small
layer of fluid near each end plate, acting to pin the fluid to the
plates. (Equivalently, this region of very high viscosity can
effectively be thought of as forming {\em part} of the plates.)  The
viscosity takes the form
\begin{equation}
\eta(u) = \eta 
\left[ 
1 + \frac{1}{32}\left( \frac{r_{\scalebox{0.6}{$0$}} e^{-\bar\epsilon}}{u} \right)^2 + \frac{1}{32}\left( \frac{r_{\scalebox{0.6}{$0$}} e^{-\bar\epsilon}}{1 - u} \right)^2 \right],
\end{equation}
in which $r_0$ is the initial  radius of the filament.

\section*{SGR model}

Full details of the SGR model in its original, spatially uniform
(zero-dimensional) form can be found in
Refs.~\cite{sollich-prl-78-2020-1997_SI,cates-jr-48-193-2004_SI}.  For a
one-dimensional adaptation to spatially non-uniform shear flows, see
Ref.~\cite{Fielding2009_SI}.  Here we detail the counterpart adaptation
to non-uniform extensional flows in a filament of soft glass.

We divide the filament equally along its length into $M$ sub elements,
called bins for convenient shorthand in what follows.  Each bin is
initialised with the same area $A_{m}$ and the same length $L_m$,
corresponding to a uniform filament initially. 

In each bin we place $N$ SGR elements.  Each element is initialised
with a yield energy $E_{mn}$ drawn randomly from the SGR model's prior
distribution of trap depths $\rho(E)=\tfrac{1}{x_{\rm
    g}}\exp(-E/x_{\rm g})$, where $x_{\rm g}$ is the glass transition temperature. In our numerics we use units in which $x_{\rm g}=1$. Each element is also assigned zero initial
strain $\epsilon_{mn}=0$ and correspondingly zero stress
$\sigma_{mn}=0$.  The tensile viscoelastic stress in any bin is 
given as $\Sigma_m=\tfrac{1}{N}\sum_n \sigma_{mn}(=0$ initially).

We then evolve the dynamics at any timestep $t\to t+Dt$ by performing
updates in three substeps, as follows.

In step 1, the elastic loading dynamics updates the strain of each SGR
element according to $\epsilon_{mn}\to\epsilon_{mn}+\edot_m Dt$, where
$\edot_m$ is the local strain rate in that element's bin. (This
$\edot_m=\edotbar$ for the first timestep. In all subsequent timesteps
it is updated in step 3 below.)  Following
Ref.~\cite{cates-jr-48-193-2004_SI}, the updated stress of each SGR
element is prescribed from its updated strain as
$\sigma_{mn}=3GQ(\epsilon_{mn})$, in which
\be
Q(\epsilon)=\frac{5}{8}\frac{1}{e^{3\epsilon}-1}\left[e^{2\epsilon}-2 e^{-\epsilon} + \left(e^{5\epsilon}-4e^{2\epsilon}\right)T\left(e^{3\epsilon}-1\right)\right],
\label{SI::Q}
\ee
with the function
\be
T(a)=\frac{\arctan(\sqrt{a})}{\sqrt{a}}.
\ee
(Note that this function $Q$ (Eqn.~\ref{SI::Q}) is distinct from the one used above for
the fluidity model (Eqn. (1) in the main text). We use this notation $Q$ here for consistency with
that of Ref.~\cite{cates-jr-48-193-2004_SI}.)  The strain of each bin is
likewise updated as $\epsilon_m\to\epsilon_m+\edot_m Dt$, the area as
$A_m\to A_m \exp(-\edot_m Dt)$ and the length as $L_m\to L_m
\exp(\edot_m Dt)$. The updated stress in each bin
$\Sigma_m=\tfrac{1}{N}\sigma_{mn}$.

In step 2, plastic relaxation is implemented by dynamics in which the
probability of any SGR element hopping during any short time interval
$Dt$ is given by $r_{mn}Dt$. Following~\cite{cates-jr-48-193-2004_SI}, we
take
\be 
r_{mn}=\frac{1}{\tau_0}\exp\left[-\frac{E_{mn}-\lambda  R(\epsilon_{mn})}{x}\right]
\ee
which defines the parameter $\lambda$. The function $R$ is given by,
\be
R(\epsilon)=\frac{1}{2}\left[e^{-\epsilon}+e^{2\epsilon}T\left(e^{3\epsilon}-1\right) \right]-1.
\ee
Once a hop has occurred for any element $n$ in bin $m$, its strain and
stress are reset as $\epsilon_{mn}\to 0$, $\sigma_{mn}\to 0$, and its
new yield energy is selected at random from the prior distribution
$\rho(E)$.

Finally in step 3, force balance along the filament is implemented by
realising that the force is the product of the area times the tensile
stress, and that this must be uniform along the filament:
\be
F_m=A_m\left[\Sigma_m +3\eta\edot_m \right] = F(t).\label{eqn:fb}
\ee
This allows us to calculate the force $F(t)$ as
\be
F(t)\frac{1}{M}\sum_m\frac{1}{A_m}=\frac{1}{M}\sum_{m}Q(\Sigma_{m})+3\eta\edotbar(t),
\ee
given a globally imposed average strain rate $\edotbar(t)$ along the
filament. This $F(t)$ is then finally fed back into Eqn.~\ref{eqn:fb}
in order to calculate the strain rate profile $\edot_m$ along the
filament.

As for the fluidity model, we choose units in which the modulus $G=1$
and the microscopic time $\tau_0=1$. We also set the glass transition
temperature $x_{\rm g}=1$.

\begin{figure}[t]
\includegraphics[width=0.40\textwidth ]{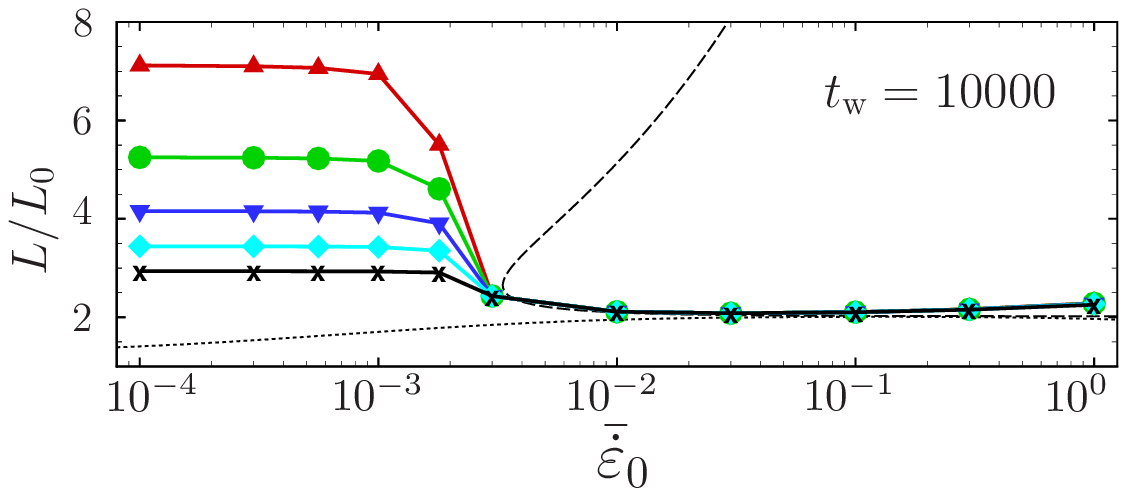}
\caption{Counterpart of Fig. 2 in the main text, for the fluidity model subjected to a constant velocity protocol (b), with a strain rate $(1/\edotbar_0+ t-\tw)^{-1}$ that progressively deceases in time from an initial value $\edotbar_0$. 
Solid lines: draw ratio at
  which the radius of the filament at its thinnest point has fallen to
  35\%, 30\% $\cdots$ 15\% of the initial filament radius (curves
  upwards). Initial aspect ratio, $\Lambda=2$. Dotted lines:
  Consid\`ere criterion.  Dashed lines: counterpart
  of the Consid\`ere criterion for elastoplastic materials.
  Parameters: $\mu=0.1$ and $\eta = 0.01$.
  \label{fig:fluidity1}}
\end{figure}
\section*{Constant drawing rate (protocol b)}

The results presented in the main text were for protocol (a) in which
a constant rate of Hencky strain $\edotbar$ is applied to the sample,
corresponding to exponentially separating sample ends
$L=L_0\exp\left[\edotbar (t-\tw)\right]$. As shown for the fluidity model in
Fig.~\ref{fig:fluidity1}, the same behaviour is seen for protocol (b),
which has a constant rate $\dot{L}$ of length increase, corresponding
to a progressively declining strain rate $\edotbar = (1/\edotbar_0+
t-\tw)^{-1}$.

\section*{Robustness to variations in parameter values}

We have checked that our results are robust to reasonable variations
in the initial aspect ratio $\Lambda=1,2,3$, and in the value of the
Newtonian viscosity $\eta$. Larger $\eta$ cause the slight upturn of
$L/L_0$ versus $\edotbar$ for old samples $\edotbar\tw > 1$ in
Figs. 2 and 3 of the main text to be
more pronounced: within this old-sample regime, more quickly drawn
samples survive to higher $L/L_0$.  Our results also apply robustly
over large regions of the space of the phenomenological parameters
$\mu$ (fluidity) and $\lambda,x$ (SGR, within its glass phase $x<1$).
For `less glassy' samples (larger $\lambda$ and larger $x$ values
approaching onset of the liquid phase at $x=1$), more complicated
effects are possible.  However the basic phenomenon that we have
reported remains robust even here: of relatively early failure onset
determined by elastic loading alone for old samples $\edotbar\tw \gg 1$,
and failure onset delayed to larger $L/L_0$ also involving plastic
relaxation for young samples $\edotbar\tw \ll 1$.

\end{document}